# Distribution of Cognitive Load in Web Search

Jacek Gwizdka, PhD
Assistant Professor
Department of Library and Information Science
School of Communication and Information
Rutgers University
4 Huntington St.
New Brunswick, NJ 08901
+1 732 932 7500 x8236
jasist@gwizdka.com
[Corresponding author]

## Abstract

The search task and the system both affect the demand on cognitive resources during information search. In some situations, the demands may become too high for a person. This article has a three-fold goal. First, it presents and critiques methods to measure cognitive load. Second, it explores the distribution of load across search task stages. Finally, it seeks to improve our understanding of factors affecting cognitive load levels in information search. To this end, a controlled Web search experiment with forty-eight participants was conducted. Interaction logs were used to segment search tasks semi-automatically into task stages. Cognitive load was assessed using a new variant of the dual-task method. Average cognitive load was found to vary by search task stages. It was significantly higher during query formulation and user description of a relevant document as compared to examining search results and viewing individual documents. Semantic information shown next to the search results lists in one of the studied interfaces was found to decrease mental demands during query formulation and examination of the search results list. These finding demonstrate that changes in dynamic cognitive load can be detected within search tasks. Dynamic assessment of cognitive load is of core interest to information science because it enriches our understanding of cognitive demands imposed on people engaged in the search process by a task and the interactive information retrieval system employed.

## Introduction

Human perception and cognition are engaged at multiple levels during information search. First, the information search process is clearly cognitive in nature (Ingwersen, 1996). Second, interaction with computing devices that mediate the search relies on human perception and cognition (Card, Newell & Moran, 1983). Web search is affected by the task, system, and individual searcher characteristics (e.g, Borgman 1986; 1989; Byström & Jarvelin, 1995; Kim, 2001; Ford, Miller & Moss, 2001; 2005; Toms, O'Brien & Mackenzie, 2008; Li & Belkin, 2008). These factors, either alone or in combination, influence the level of difficulty experienced by a searcher. One kind of difficulty is related to mental, i.e. cognitive, requirements imposed by the search system or the task itself.

Understanding what contributes to a user's cognitive load during search tasks is crucial to understanding the search process and to identifying which search tasks types and search system features make greater demands of users. We also need to better understand how novel user interfaces and interactive functionality affect user performance and system usability, usefulness, and acceptance. For example, in one system a user relevance feedback feature was avoided by users due to the heightened cognitive load (Back & Oppenheim, 2001). Detecting mental effort levels experienced by a user during a search process is important for understanding cognitive demands imposed by search tasks, user interfaces, and information displays, and for identifying where, and possibly how, to lower the mental effort required for effective interaction. For these reasons, the dynamic aspects of cognitive load are of core interest to information science.

The results presented in this article show that cognitive load differs between search task stages and that different components of cognitive load (e.g., intensity, peak load) tend to be related to different aspects of search task performance. A variation of the dual-task methodology is used to show how cognitive load is sensitive to the dynamic changes in task demands such as the changes of load from one task stage to another.

The article is structured as follows. In the next section, the concept of cognitive load is presented. Next, cognitive load assessment techniques are described and selected information science research that examined cognitive load is discussed. Emphasis is placed on the manifold nature of cognitive load and its dynamic properties, and their implications for assessing cognitive load in information science. Finally, the study, its results and a discussion of those results are presented.

# Background and Related Work

This section discusses the concept of cognitive load and its components. It then presents an overview of measurement techniques, and discusses types of cognitive load.

## The Concept of Cognitive Load

Cognitive load is used to evaluate, model and predict human performance in several disciplines, including cognitive and educational psychology, human factors and engineering psychology. The complexity of the cognitive load concept is reflected in the disciplinary differences in its understanding. Even within a single discipline, there may be no common definition of cognitive load. In human factors and engineering psychology cognitive load is referred to as mental workload. In one view, it is described as the portion of a person's processing capacity actually required to perform a particular task (O'Donnell & Eggemeier, 1986). In this view, the load is an external variable, which is a task characteristic (Hancock & Chignell, 1987; Tsang, 2002). In another view, mental workload is described in terms of an interaction between task requirements and human capabilities or resources (Hancock & Chignell, 1987; Tsang, 2002). Here, workload is understood as the relation between the demand for mental resources imposed by a task and the person's ability to supply those resources (Moray 1979). Thus, the load is relative to the user and the task being completed in a given environment. Educational psychologists generally share this perspective.  They explicitly consider several distinct sources of cognitive load and specify them in the Cognitive Load Theory, which describes three types of cognitive load (Chandler &

Sweller, 1991; Sweller, van Merrienboer, & Paas, 1998). *Intrinsic load* is the load imposed by the problem being solved (task source). *Extraneous load* is the load imposed by the environment where the task is being performed (external source). *Germane load* represents intentional effort invested by an individual in learning that goes beyond simple cognitive task performance and comprehension (Schnotz & Kürschner, 2007). The learning effort involves committing information from working memory to long-term memory and is therefore related to long-term changes in an individual's mind.

The concept of cognitive load is closely related to the notion of limited mental resources. The Multiple Resource Theory (Wickens, 2002) says that humans have a finite set of mental resources of several types. When demands of one task are high, the resources committed to that task become unavailable to a second task if it requires the same type of mental resources (e.g., visual vs. auditory) and at the same stage of processing (e.g., cognitive vs. response-related). The mental limitation in perception and cognition are a result of limited capacity of working memory. Working memory is conceptualized as containing three subsystems, one responsible for verbal information processing (phonological loop), one for visual information processing (visuo-spatial sketchpad) and one for controlling and coordinating the processing machinery (central executive) (Baddley & Hitch, 1980)[1]. Sensory information and information recalled from long-term memory passes through and is processed in working memory. A limitation in the number of "slots" in working memory is used to explain some constraints on human capacity for cognitive processing. This limited capacity was first described by George Miller in his famous paper on "the magic number seven plus or minus two" (Miller, 1956). Newer findings suggest the actual limit of working memory capacity may be lower than the one proposed by Miller. For example, Cowan (2009) postulates that the limit is four items and Oberauer (2002) suggests it is only one item. Others argue the limitation on human mental capacity is not in the limited number of memory "slots", but in the control of attention that is required for effective processing of information (Engle, 2002; Kane & Engle, 2000). Notwithstanding these disputes, the fundamental notion of limited mental capacity is commonly agreed upon.

## Cognitive Load Measurement

A common classification of cognitive load assessment techniques divides them into *performance, subjective* and *physiological* measures. For a detailed review the reader is referred to Cegarra & Chevalier (2008). An alternative classification uses four dimensions: *analytical–empirical, objective–subjective, direct-indirect,* and *dynamic–static*. In this paper we will use the common classification system, but annotate the measurement methods with attributes that place them along some of the four dimensions in the alternative system. Our focus will be on *empirical* methods. A separate group of cognitive load assessment methods is *analytic*. These methods include expert opinion, task analysis, simulation and interface inspection based on models of human perception and cognition (e.g., Card, Newell & Moran, 1983; Polson, Lewis, Rieman & Wharton, 1992; John & Kieras, 1996; Teo & John, 2008). We will mention two specific analytic approaches in the section "Cognitive Load on Information Tasks".

*Subjective methods* include self-rated scales, think-aloud protocols, and post-task interviews. One of the best-known and most widely used instruments is the NASA Task Load indeX (TLX) (Hart & Staveland, 1988). Subjective measures can shed light on the cognitive state of the user and are important for assessing user's perceptions of the task. These methods are expected to measure overall load (see section on Cognitive Load Types). Subjective measurements are typically taken at one point after a task is completed and are thus *static*. The static nature of these methods makes them inappropriate for assessing dynamic changes in cognitive load that are focus of this article.

---

[1] Newer versions of Baddley's model contain a fourth component, "Episodic Buffer", responsible for linking information across different senses and fixing chronological order.

*Performance measures* include performance on the *primary task* and performance on the *secondary task*. Performance measures on the primary task are *objective*. They include number of errors, accuracy, task completion time relative to user population time, and ratio of the actual completion time to an ideal completion time (Wickens, 2002). Primary task performance measures seem to be particularly applicable when task performance pace is externally controlled (e.g., by external events, such as factory parts delivered to a worker for assembly). When task pace is controlled by a task performer, applicability of these measures is uncertain. In an information search task, a user may take longer to complete the task for many reasons other than higher demand on mental resources.

Methods that involve performance on a *secondary task* are called *dual-task techniques* (Brünken, Steinbacher, Plass & Leutner, 2002; Kim & Rieh, 2005; Cegarra & Chevalier, 2008). They are *objective* and *direct* (Brünken, Plass & Leutner, 2003). Cognitive load measures derived from performance on secondary task are grounded in the notion of limited cognitive resources and, in particular, in Multiple Resource Theory (Wickens, 2002). A secondary task should be designed to require the same metal resources as the main task. If two (quasi-) simultaneously performed tasks require the same resources (e.g., verbal working memory), then the available resources need to be distributed between the tasks. This has the effect of linking the performance on the task. The effect on secondary task performance is expected to increase as more resources are required by the primary task. Secondary tasks typically involve some aspects of monitoring external events, which are periodically delivered through the visual or auditory channel. The choice of sensory channel for a secondary task depends on the nature of a primary task. The standard dual-task metrics for assessing cognitive load are reaction time to secondary task events and the number of misses of secondary task events. The frequency of secondary task events should be selected such that it is large enough to affect performance on the primary task but still allows for close-to-normal performance.

*Physiological measurement* techniques are *objective* and include electro-encephalography (EEG), functional near infrared spectroscopy (fNIRS) (Hirshfield, et al., 2009), and eye-tracking. In particular, changes in pupil diameter have been used to assess mental load in studies in information science and human-computer interaction (e.g., Iqbal, Zheng & Bailey, 2004; Tungare & Pérez-Quinones, 2009).

Dual-task and physiological measurement techniques have the advantage of enabling *dynamic*, real-time data collection during task. However, interpretation of data collected by physiological techniques can be challenging, and the use of external devices can also be expensive and impractical. In contrast, the dual-task method is attractive because it allows for an inexpensive and objective assessment of effort on the primary task[2]. Only a few studies in information

---

[2] The typical dual task method has a limitation of a controlled lab setting. However, one could also imagine using it

science have employed the dual-task method to assess cognitive load in online search tasks (e.g., Kim & Rieh, 2005; Dennis, Bruza & McArthur, 2002). We present results from these and other studies in the next section.

## Cognitive Load on Information Tasks

Several interactive information system studies have measures cognitive load. In this section we describe the findings and measurement methodologies that were employed.

We start with work that takes an analytical approach and considers visual complexity in information displays. One expects that as the complexity of visual information representations increases, the user's cognitive load increases because of increased demands on perceptual and cognitive processing resources. Rosenholtz, Li, & Nakano (2007) define visual clutter as "the state in which excess items, or their representation or organization, lead to a degradation of performance at some task" (p. 3). Since consideration of all possible interaction tasks with information displays would be daunting, Rosenholtz et al. identify visual search as a low-level task that is common to many higher-level information display interactions. Predictions using their Statistical Saliency Model were compared with the users' reaction times on visual target search tasks. Images with higher congestion of visual features were found to impose higher processing demands on the users.

The ViCRAM[3] project takes an empirically grounded approach to establishing analytical models of visual and cognitive complexity. Harper, Michalidiou & Stevens (2009) examined users' subjective perceptions of web page complexity. Participants were asked to rate the visual complexity of diverse Web pages and then to describe the ratings. Participants tended to describe the visual complexity of a Web page by referring to objects that had greater interaction complexity. For example, Web pages that contained visually simple Web forms were rated as more complex than pages that contained visually complex images. Harper et al. concluded that visual complexity of a Web page could be used as an implicit measure of cognitive load required for user processing and interaction with the page. They suggested cognitive load could be estimated analytically without an explicit measurement of user performance on a task. In contrast to Rosenholtz et al., such estimation should incorporate features requiring interaction and not be derived solely from an analysis of visual features. These conclusions have been confirmed, in part, in a follow up study (Michalidiou, 2009).

Gwizdka and Spence (2007) used subjectively-assessed visual complexity of web pages in combination with the length of a navigation path on a website to assess cognitive task difficulty. One measure of the user interaction with the website was the user's navigation path. The navigation path measure was quite effective; it was found to be positively correlated with the searcher's effort (number of pages visited) and the subjective assessment of task difficulty.

Wästlund, Norlander & Archer (2008) studied the effects of display text presentation on mental workload. Text documents that fit on a screen and could be navigated by screen-full chunks were

---

in a natural setting. For example, the secondary task could be designed as an email or instant messaging notification.
[3] ViCRAM stands for Visual Complexity Rankings and Accessibility Metrics (http://hcw.cs.manchester.ac.uk/research/vicram/).

compared with "continuous" text documents that could be navigated by scrolling. Mental effort was assessed using a dual-task method, where an average reaction time to the secondary task was calculated per task. The primary task consisted of reading a document, and so the results may be related to one stage of a search task. Stress and tiredness were subjectively assessed before and after each task. Wästlund et al. found that mental workload was lower when the text was fitted to the screen and did not have to be scrolled and concluded that optimized document layout is needed to minimize mental workload.

Schmutz, Heinz, Métrailler & Opwis (2009) studied ecommerce tasks that involved finding five predefined books in four different online bookstores. The tasks were performed by browsing product hierarchies; without using the sites' search engines. Cognitive load was assessed by measuring performance on a secondary task and by using subjective assessment after each task (NASA TLX). The secondary task involved visual monitoring of changes in color (from green to red) of a single letter displayed on the right side of the screen. Two secondary-task measures were obtained: reaction time to the changes in color and the miss-rate. Although both tasks and sites were controlled to some extent, the focus was on the levels of mental load imposed by different interfaces. The analysis was performed for each bookstore by averaging the collected data across tasks. Schmutz et al. found that the subjective load ratings differed between the online bookstores and were correlated with task completion times (positively) and with user satisfaction (negatively). There was also a weak correlation between the miss-rate and the subjective load ratings. However, the secondary task reaction times did not differ between the sites. The correlation between subjective load and other measures (time on task and subjective satisfaction) may indicate that the subjective load ratings measure an overall performance effort and not just cognitive load. The lack of differences between reaction times (RT) across the interface conditions may indicate that RT and the subjective method measure different aspects of cognitive load. An alternative explanation is simply that primary task effects on secondary task load exist but average to zero as interaction time increases. The next study provides support for this explanation.

Iqbal, Zheng & Bailey (2004) used changes in pupil size to assess cognitive load on four tasks of different types (reading comprehension, information search, mathematical reasoning, and object manipulation). They found no effects when load was averaged at the task level, however when tasks were re-examined at a sub-task level using task decomposition, they found significant differences in the levels of cognitive load.

Kim & Rieh (2005) used a dual-task method to measure cognitive load during web search. They found user evaluation of search result lists and documents (content) had different cognitive loads as measured by the number of misses of the secondary task event. Users experienced lower loads when viewing search results, and higher loads when reading documents. The authors explained this difference by assuming that subjects tended to skim search results but read documents. Their result corroborates the Iqbal, Zheng & Bailey (2004) finding concerning differences in cognitive load levels between sub-tasks and search task stages.

Dennis and colleagues conducted several studies that employed dual-task methods (Dennis, McArthur & Bruza, 1998; Dennis, Bruza, & McArthur, 2002). Their 1998 study used the Excite search engine and found lower cognitive load (lower number of the secondary task misses) during the query refinement state as compared to the search result list state (the 'document summaries' state). Dennis et al. (2002) describe two studies that used Yahoo directory, Google, and Hyper Index Browser. The Hyper Index Browser allowed the user to view the underlying document space at a higher level of abstraction and supported navigation by offering query refinements derived from the document space. They found no effects in the second study reported in the 2002 paper, hence, we will not discuss it further. In the first 2002 study, the authors found that cognitive load was lower (faster reaction times to the secondary task, and no difference in terms of misses) when perusing the Google search engine results list page than when perusing query refinements in the Hyper Index Browser. The result obtained in the first 2002 study was in the opposite direction compared with the result of the 1998 study. The reversal of the cognitive load effect was explained as an effect of differences in the search result display and in the query refinement navigation. Google had a lower load than Excite because Google presented the document result list with respect to query terms, while Excite did not present the list of results in relation to query terms. The Hyper Index Browser supported a view of the underlying document space at a higher level of abstraction and allowed for navigation through query refinements. Making sense of the higher level of abstraction apparently required more mental effort. Lower familiarity of users with the Hyper Index Browser and its associated query-by-navigation metaphor may have also affected measured cognitive load.

Gwizdka (2009) performed initial, task-level cognitive load analysis of the data collected in the study presented in this article[4]. Subjective ratings of task difficulty differed between the tasks in the expected direction, however, the two standard secondary-task measures of cognitive load (reaction time and missed events) did not significantly differ at the task level. Since this result seemed surprising, additional measures were created: a ratio of correct responses to all responses to the secondary task and a ratio of user estimated secondary task events to the actual number of the secondary task events. These two measures showed significant, but weak, differences between the tasks. The ratio of correct responses was 94% for tasks of low and medium subjective difficulty compared with 87% for difficult tasks. The number of secondary task events was found to be over-estimated by 30% for easy tasks, while no difference was found for medium and difficult tasks. Over 50% of cases search tasks were assessed as easy, so that may explain the results. This could be also due to the studied user population or the tasks employed in the study (see "User Tasks" section). A more interesting explanation is suggested by related work presented earlier. We know that one can expect differences in cognitive load between task stages (Dennis et al., 2002; Kim & Rieh, 2005), and that when differences are absent at the task level, they may exist at the task-stage level (Iqbal, Zheng & Bailey, 2004). These observations motivate, in part, the analysis presented in this article.

Summarizing, we note that when cognitive load was assessed using dynamic measures, the results depended on the unit of analysis (the unit of averaging). Generally, when dynamic measurements were averaged at the task level, no differences were found. In one case, where task-level differences were found the tasks were narrowly defined and limited to one kind of activity (Wästlund et al. 2008). When cognitive load was averaged at the sub-task (task-stage) level, the differences were typically found. Discovery of differences in cognitive load for different task stages should not be surprising in view of previous research that examined complexity and the associated cognitive demands at the level of information display (e.g.,

---

[4] This article, in contrast, focuses on analysis at the task-stage level.

Page 7

Rosenholtz et al. and Harper et al., presented in this section), and given our understanding of differences between task-stages (Dennis et al., 2002; Kim & Rieh, 2005). More generally, elements of search tactics and strategies involved in interactive information search (e.g, Bates, 1979; Belkin, Cool, Stein & Thiel, 1995; Marchionini, 1995; Bhavnani & Bates, 2002; Xie, 2002) imply differences in cognitive effort as different tactics and strategies are employed, for example query reformulation vs. monitoring content for changes. It may be surprising, though, that relatively little attention has been given to the study of cognitive load at units shorter than a task. The next section places cognitive load measurement techniques and studies in a wider perspective by providing a framework for cognitive load types.

## Cognitive Load Types

As we have seen, different cognitive load assessment techniques are often adopted by researchers without an explicit consideration of what aspect of load they measure. An assumption seems to be made that cognitive load is just one construct. One specific aspect that seems to be all-too-often overlooked is the dynamic characteristic of cognitive load changes during task performance. These changes reflect shifts in the demands on the user from both the task and the interactive system. Even when dynamic measurement methods are used, cognitive load has usually been averaged over an entire experimental condition (a task or a user interface). The nature of dependencies between cognitive resources and human performance described in the Multiple Resource Theory (Wickens, 2002) is dynamic and not based on average values. Hence, the averaging may mask the dynamic aspects of the load. It is useful, then, to consider frameworks that reflect all of the aspects of cognitive load in order to better understand the relationship between cognitive load measures and their use and analysis under alternative experimental conditions.

Xie & Salvendy (2000) proposed a comprehensive framework of cognitive load that helps to understand the variety of cognitive load aspects and their measurement. The framework outlines cognitive load constructs and provides a systematic way of thinking about the types of cognitive load. The framework includes five (four objective and one subjective) load types: 1) instantaneous load is the load at a specific moment. It is the basic measure from which other measures are derived; 2) peak load is the maximum value of the instantaneous load in a given period of time; 3) average load is intensity of load measured per unit of time; 4) accumulated load is the whole load experienced up to any point in task performance; and 5) overall load is a subjective mapping of instantaneous, average, and accumulated workloads in a person's mind and is expected to be correlated with both the average and the accumulated load. Instantaneous load is used to derive the other load measures, so the framework is grounded in the temporal dimension of cognitive load. Researchers in other areas have pointed out the need to pay attention to the dynamics of cognitive load. For example, van Gog, Kester, Nievelstein, Giesbers & Paas (2009) observed (p. 328): "*It would be interesting [...] to study how processing demands [...] fluctuate in different phases of the learning process and how these patterns are related to overall cognitive load and investment of mental effort*." Most existing techniques measure one kind of cognitive load. Subjective measures assess the overall load, whereas physiological and performance measures can be used to assess the other load types. In particular, the dual-task technique measures the instantaneous load, but it is typically used to calculate the average load over some unit of user interaction.

# Research Motivation and Objectives

Our discussion of cognitive load types and their measurement makes it clear that different assessment methods tap into different aspects of cognitive load. Outcomes of different measurement methods thus should be considered as complementing each other and, in most cases, should not be directly compared. The discussion also points to the sensitivity of selecting an appropriate unit, for example, the span of a task vs. a task phase, over which to average the values obtained from the dynamic assessment of cognitive load. Averaging over longer units may miss fluctuations in load levels and, in particular, peaks in load. Making the unit shorter will better reflect the dynamic changes in the level of cognitive load. These observations, together with our initial, task-level analysis (Gwizdka, 2009) provide motivation for the research presented in this paper.

Based on the foregoing analysis, this paper focuses on assessment the peak load and the average cognitive load at the task-stage level. Calculating average load at this finer granularity will facilitate understanding of the dynamics of cognitive load patterns. Calculating peak load will identify situations where a person's cognitive capacity may have been exceeded. If the peak load exceeds a person's capacity, the performance on a task may be degraded and perhaps even completely inhibit their ability to complete the task.

We expect to find differences between cognitive load levels calculated at the task-stage level, and, possibly also find differences in the levels of peak cognitive load between task-stages. Search task stages differ in terms of the amount of information externally provided to a searcher (e.g., displayed). Task stages such as query formulation or user description of a relevant document typically provide less external support as compared to task stages such as examining search results and individual documents. The availability of externally provided information makes the latter group of task stages take advantage of recognition processes rather than recall, while the former task stages rely more on recall. Recall from memory is in most cases more difficult than recognition (Eagle & Leiter; 1964; Norman, 1988; Baddeley, 1999). One can thus hypothesize that task stages from the first group put more demands on working memory and involve more recall from long-term memory as compared to task stages from the second group. Hence, cognitive load required by first group of task stages can be expected to be measurably higher than by the second group. Since the visual and interaction complexity required by an interface is related to cognitive load (Harper et al., 2009), one can hypothesize that there may also be a user interface effect on cognitive load demands. A more complex interface can be expected to impose a measurably higher cognitive load on a user.

More generally, this paper aims to draw attention to the study of dynamic changes of cognitive load and their importance for information science. Knowledge of cognitive load dynamics will help us to learn more about search processes at the sub-task level, about differences in mental demands between search moves, tactics, and strategies and about the effects of user interface on the searcher's mental effort.

# Method

## Participants

Forty-eight subjects (Table 1) participated in a question-driven, web-based information search study conducted in a controlled experimental setting. Participants were Rutgers University students, mainly from undergraduate and graduate programs offered by the School of Communication and Information (Table 1). Participants were offered a monetary incentive ($20); participants who were recruited from an undergraduate HCI class received a partial course credit. Participants were motivated by a monetary performance-based bonus of the same value as the monetary incentive ($20). The bonus was administered to one third of the participants after the experiment was completed and the task outcomes were calculated. Most participants were frequent Web searchers and only one person searched the Web relatively infrequently: once or twice a week (Table 1).

Table 1. Participants' profile

| | |
|---|---|
| **Age** | Mean 27 year; median 23 years. range 20-51 |
| **Gender** | 17 females and 31 males. |
| **Current level of study** | 65% - undergraduate (48% from Human-Computer Interaction class in the Information Technology & Informatics program; 5% from other Information Technology & Informatics classes; 12% from Communication, Journalism and Media Studies); 6% - Master (4% MLIS); 23% - PhD (from 10 different programs ranging from sciences, engineering to humanities); 6% - other (just graduated) |
| **English language** | First language 56%; spoken at home 65% |
| **Web search frequency** | 35% almost constantly; 46% several times a day; 17% once a day; 2% once or twice a week. |

## Procedure

Each study session took an hour and a half to two hours and was conducted in a university lab on a personal desktop computer running the Microsoft Windows XP operating system. Each session consisted of the following steps: an introduction to the study, consent form, search task practice, background questionnaire, six search tasks, and post-session questionnaire. Before and after each search task, participants answered a short set of questions to provide subjective assessments of task characteristics (before) and their performance on a task (after). The searchers bookmarked and tagged the web pages that they considered relevant. User interaction with the computer (visited and bookmarked URLs, mouse and keyboard events, and video from a screen cam) was recorded using Morae software[5]. The start and end of each search task were controlled by an external program that was used to start and end a Web browser session (Internet Explorer). The same program was used to log task start and end times. Total time on a task included the searcher's bookmarking activity.

## User Tasks

The study search tasks were designed as questions that described what information needed to be found. The tasks were designed to differ by difficulty and structure. A total of twelve questions were used in the study. Four tasks were created by us, while eight were created by Toms et al. (2008). Two types of search tasks were used: Fact Finding (FF) and Information Gathering (IG) (Kellar, Watters & Shepherd, 2007). The goal of a fact finding task, also called a known item search (Li & Belkin, 2008), is to find one or more specific pieces of information (e.g., name of a person or an organization, product information, a numerical value; a date). The goal of an information gathering task, also called a topical search, is to collect several pieces of information about a given topic. The tasks were also divided into three categories according to the *structure* of the underlying information need (Toms et al., 2008), 1) Simple (S), where the information need is satisfied by a single, independent piece of information (by definition, simple task is of the fact finding type); 2) Hierarchical (H), where the information need is satisfied by finding multiple characteristics of a single concept; this is a depth search, where a single topic is explored; 3) Parallel (P), where the information need is satisfied by finding multiple concepts that exist at the same level in a conceptual hierarchy; this is a breadth search. Task types and structures are listed in Table 2. An example task of each type is provided in Appendix A. The tasks were constructed using Situated Work Task Situations (Borlund, 2003). The simulated situations were created by using task scenarios that provided participants with the search context and the basis for relevance judgments.

---

[5] Morae is a product of Tech Smith Inc. http://www.techsmith.com/morae.asp

Based on their characteristics, the tasks were categorized into three levels of *objective difficulty*. FF-S tasks were assigned a low difficulty level; FF-P and FF-H tasks a middle-difficulty level; and IG-H and IG-P tasks a high difficulty level. The three levels of objective difficultly were assigned based on the quantity of desired outcomes (single or multiple) (Campbell, 1988; Bell & Ruthven, 2004; Li & Belkin, 2008), and on the degree of a priori determinability of a task (Byström & Jarvelin, 1995).

Table 2. Task Types and Structures. (FF-Fact Finding, IG-Information Gathering)

| Task Acronym | Task Structure and Type | Quantity of Outcomes | A priori Task Determinability | Objective Difficulty |
|---|---|---|---|---|
| **FF-S** | Simple fact finding task (known item search) | single | high | Low |
| **FF-P** | Parallel fact finding task (known item search) | multiple | high | Medium |
| **FF-H** | Hierarchical fact finding task (known item search) | multiple | high | Medium |
| **IG-P** | Simple information gathering task (topical search) | multiple | low | High |
| **IG-H** | Parallel information gathering task (topical search) | multiple | low | High |

During the course of an individual study session, each participant performed six tasks of differing type and structure (Table 3). Thus, the forty-eight participants performed a total of 288 tasks. For each task, the participant was able to choose between two questions of the same type and structure but on different topics. We offered the choice of topics to increase the likelihood of a participant's interest in the question topic. The order of tasks was balanced with respect to the objective task difficulty. Two task orders were repeated: cases where the difficulty increased from low to high, and cases where the difficulty decreased from high to low (Table 3). We repeated only these two task orders, because earlier research (Gwizdka & Spence, 2006) showed we could assume that task ordering where the difficulty is not monotonically changing (e.g., medium-low-high, high-low-medium) is not likely to result in significant effects. The simple fact-finding task was repeated twice (e.g, FF-S) in each rotation. To avoid repeating search topics in one task rotation, we created two groups of search topics for this task type (for the total of four).

Table 3. Task rotations (for one rotation of each search system). Numerical indexes refer to FF-S topic groups.

| QR / Task Seq. | TSeq1 | TSeq2 | TSeq3 | TSeq4 | TSeq5 | TSeq6 |
|---|---|---|---|---|---|---|
| QR1 | FF-S1 | FF-P | IG-H | FF-S2 | FF-H | IG-P |
| QR2 | IG-H | FF-P | FF-S1 | IG-P | FF-H | FF-S2 |
| QR3 | FF-S1 | FF-P | IG-H | IG-P | FF-H | FF-S2 |
| QR4 | IG-H | FF-P | FF-S1 | FF-S2 | FF-H | IG-P |

## User Interfaces

The search tasks were performed on the English Wikipedia by using two different search engines: UI1 - Google Wikipedia search (Figure 1), and UI2 - ALVIS Wikipedia search (Buntine et. al. 2005) (Figure 2). Wikipedia's own search engine was not used. UI1 was familiar to study participants, while UI2 was not. UI1 displayed search results in a list. UI2 displayed search results in a list and added categories to them. The categories were shown on the left side of the list, as well as beneath each result. Applying the analytical framework from the ViCRAM project (Harper, Michalidiou & Stevens, 2009), we can expect that the addition of categories to search results in Alvis will lead to increased complexity of user interactions because there are more interactions possible and more choices to be made in Alvis than in Google. The displayed categories were clickable and could be used to narrow down search results. These features in Alvis were expected to demand more mental effort because of the increased extraneous cognitive load of the results list display. The search engine (along with the associated interface) was switched after task 3. The four task rotations (Table 3) were repeated for two orders of user interfaces (UI1/UI2 and UI2/UI1). Thus there were a total of eight combined rotations of tasks and UIs for each participant.

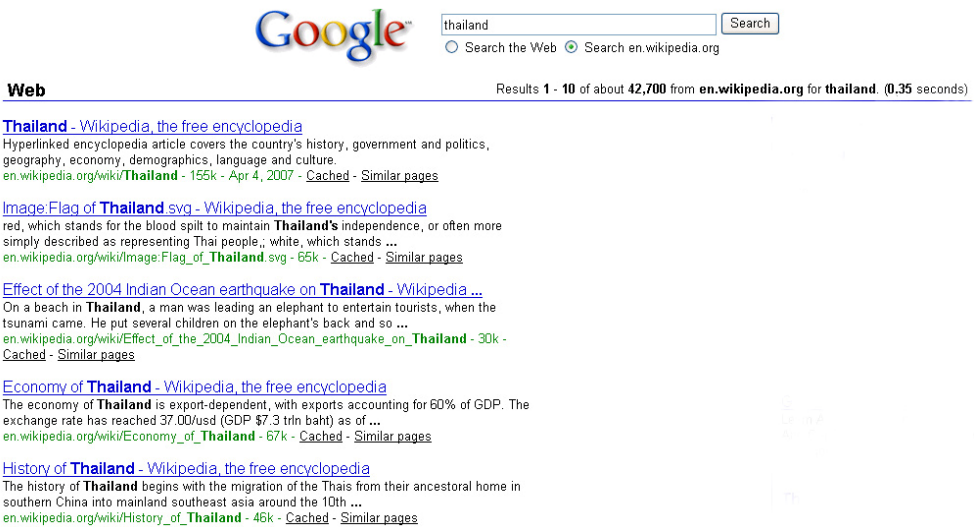

Figure 1. User Interface 1 – Google search (on Wikipedia).

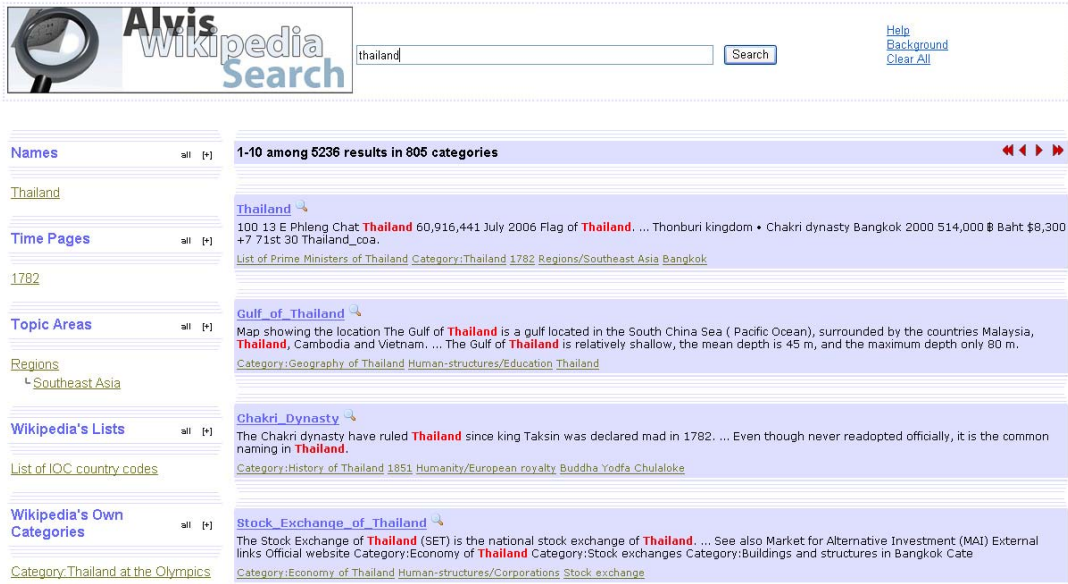
Figure 2. User Interface 2 – Alvis Wikipedia Search (Buntine et al. 2005)

## Secondary Task

A secondary task (DT), based on the Stroop effect (Stroop, 1935), was introduced to obtain indirect objective measures of user's cognitive load on the primary task. The Stroop effect demonstrates interference between attentional processes that affects reaction time to external stimuli. In the classic Stroop task, words that name colors are displayed to a participant in colored ink. The participant is asked to name the ink's color ignoring the word. When a word's ink color differs from the word's meaning (e.g., the word "yellow" printed in green ink), naming the color takes longer than when the meaning of the word matches the ink color.

In our implementation of the secondary task, a small pop-up window was displayed at a fixed location on the displayat random time intervals (15- 29 seconds) and for a random period of time (5-9 seconds). The length of a cycle was thus between 20 and 38 seconds. The times were based on pilot experimental sessions and previous dual-task work (Wästlund et al., 2008; Schmutz et al., 2009) and selected to be short enough to affect performance on the search tasks but still allow for close-to-normal performance. The pop-up contained a word with a name of a color (Figure 3) and a color font that matched the name in some cases. Participants were asked to click on the pop-up as soon as they noticed it. The click was performed either on the right (match between the color name and the font color) or on the left mouse button (no-color-match). The pop-up window disappeared after a random period of time or as soon as it was clicked. In case of a color match, the cognitive load imposed on a person is less related to semantic processing than in case of a no-color-match, when the cognitive load imposed on a person is higher due to more semantic processing demanded (Kane & Engle, 2003). In both cases, the motor effort required to respond to the secondary task is similar. Performance on the two modes of the Stroop task can be considered separately. Our secondary task involved motor action, as well as verbal/semantic processing (phonological loop in working memory; Baddley, 1986). The modalities of the primary task and the secondary task overlapped, and it is reasonable to assume different levels of cognitive effort on the primary task should be reflected in the differences of performance on the secondary task, and in particular, in the reaction time to the secondary task. To our knowledge, Stroop task has not been previously used as the secondary task in studies of interactive systems.

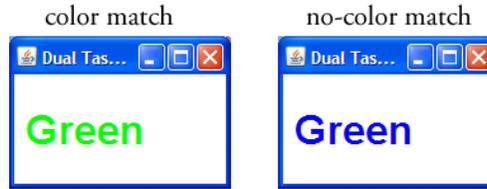

Figure 3. Secondary task pop-up window. An example of color match on the left and an example of no color match on the right.

## Independent Factors (IF)

### Task and User Interface Factors

The controlled factors included the search task characteristics and the search system interface (Table 4).

Table 4. Independent factors (IF).

| Variable Group | Variable Name | Variable Description and Scale |
|---|---|---|
| Task Characteristics | Obj_Task_Diff | Objective Task Difficulty: low-medium-high (see Table 2) |
| Search System Interface | UI | UI1Search system: UI1=Google's Wikipedia search, UI2=ALVIS Wikipedia search |

### User Characteristics

Participants were tested for two cognitive abilities (operation span and mental rotation). The tests were administered on the same computer that was used for searched (Table 5). The particular cognitive factors were selected because they were likely to affect searchers' task performance on tasks (Chen et al., 2000; Chen et al., 2004; Gwizdka & Chignell, 2004).

Table 5. Tests of cognitive characteristics of users.

| Cognitive Factor | Measures | Test Name | Short Description | Reference |
|---|---|---|---|---|
| Operation Span (WM) | OpSpan ratio. 0-100%. Higher score → higher ability | CogLab on CD (Wadsworth) | Operation Span is one of the measures of working memory (WM) performance. Operation span predicts verbal abilities and reading comprehension. | (Francis & Neath, 2003) |
| Mental Rotation (SA) | Mental Rotation Mean Reaction Time (RT) and Mental Rotation ratio of correct responses. | PsychExperiments (Dept. of Psychology, University of Mississippi) | Mental Rotation is the ability to mentally manipulate spatial images (spatial ability – SA). | (McGraw et al., 1999; McGraw & Tsai, 1993; Shepard & Metzler, 1971) |

The participants were asked about their dominant handedness and their constraints on perceptual and cognitive processing. Five out of forty-eight participants (10%) were left-handed. The participants were allowed to position mouse and the keyboard according to their preferences. The small number of left-handed participants did not significantly affect the cognitive load results. One participant reported a minor dyslexia, and one reported a minor color blindness.

## Dependent Variables

### Cognitive load measurement (CL)
User performance on the secondary task (pop-up window) was used to obtain indirect objective assessment of the user's cognitive load on the primary task. Two measures are typically calculated when using dual-task methods: a) the rate of missed secondary task events and b) reaction time to the secondary task. We calculated both of these measures and found reaction time to be a more sensitive measure than the miss rate. The measured discrete values of reaction times were used to assess average cognitive load over a given period of time (duration of a task stage) and to assess the peak load during a period of time. We performed these calculations for all secondary task events ($RT_{all}$), and also for color match ($RT_{match}$) and color non-match ($RT_{nomatch}$) cases. The resulting three variables were used as separate dependent variables.

### Subjective measures (SU)
Subjective variables were those that were self-reported by participants. Our prior publication (Gwizdka & Lopatovska, 2009) focused on the analysis of subjective variables in their relation to objective measures of the searchers' behavior at the search task level. In contrast, this paper focuses on cognitive load at the task-stage level. We will only refer to three subjective variables: pre-task assessment of familiarity with and interest in the search task topic, and post-task assessment of the task difficulty (Table 6).

Table 6. Subjective variables (SU).

| Variable Group | Variable Name | Variable Scale |
|---|---|---|
| **Before Each Search (SU-before)** | before_interest | 1=not at all / 5=extremely interested |
| | before_familiarity | 1=not at all / 5=extremely familiar |
| **After Each Search (SU-after)** | after_easy | 1=very difficult / 5=very easy |

## Task Stage Segmentation
This section presents the procedure that was used to label behavioral data (interaction logs) with task segment names. The task segments were based on selected sub-processes from Marchionini's information seeking process (Marchionini, 1995) (Figure 4).

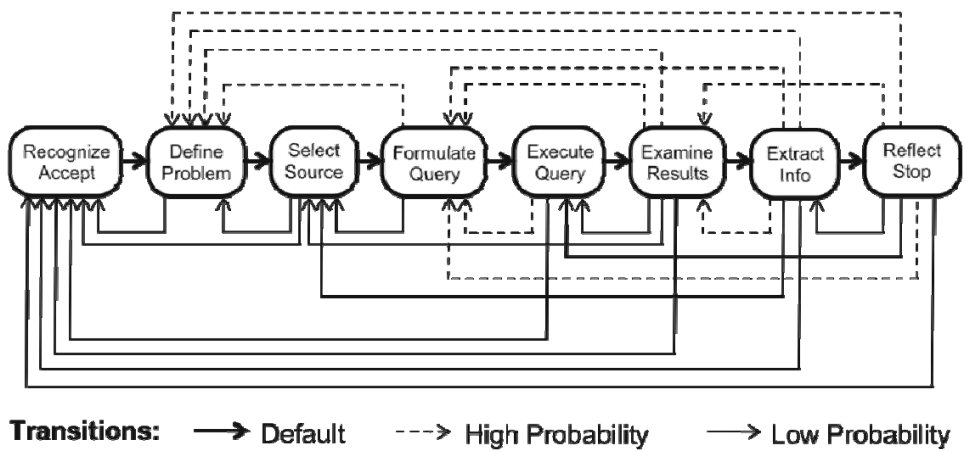
Figure 4. Information Seeking Process, (ISP) Marchionini (1995) [figure from the original source]

Task segment identification was based on the observable physical actions logged during search sessions and on the type of web page visited. Web page URLs were parsed and classified into four categories: 1) search engine home, 2) search results list page, 3) individual result page (content page), 4) bookmarking and tagging page (Table 7). Unknown URLs were generally categorized as other content pages. Out of the total 970 visits to content pages, 57 (5.9%) were visits to unique content pages. A small number of other pages (13) were identified as system pages (for example, the bookmarking system's login page and the online questionnaire pages) that were mis-classified as belonging to search episodes.

Table 7. Number of unique pages of different types visited in the study.

| Page Type | Number | Percent |
| --- | --- | --- |
| Search engine home | 2 | 0.1% |
| Search results list | 1128 | 42.1% |
| Content (Wikipedia) | 923 | 34.1% |
| Content (unknown) | 57 | 2.1% |
| Bookmark and tag | 563 | 21% |
| Other (system) | 13 | 0.5% |
| Total | 2681 | 100% |

This approach assumed the captured interaction log was sufficient for identifying task stages. In some cases, the temporal boundaries of task stages were well defined (e.g., when a task stage was entered on a new web page, such as search results list), in other cases the observed physical actions provided only an indirect glimpse into the searcher's cognitive processes. This limitation applies mainly to segmenting the query formulation stage, as that was based on keyboard activity. Table 8 presents the task stages. Task stage "Search engine home" (H) will not be further considered, since the source and the starting search engine page was given to users.

Table 8. Task stages and Marchionini's information seeking processes.

| Task stage code | Task stage name | ISP Sub-process | Description |
| --- | --- | --- | --- |
| H | Search engine home | between: *Select Source* and *Formulate Query* | Initial stage, before the first query is formulated. In this study, the source was selected for users. |
| Q | Query formulation | *Formulate Query* | Search query is formulated or re-formulated. |
| L M | Examination of search result list (L-first result page, M-subsequent result pages) | *Examine Results* | In a response to entered query, a search engine results list is displayed and examined. |
| C | Examination of an individual result (content) | *Extract Information* | Individual result (content page) is visited and is visually scanned or read. |
| B | Bookmarking and tagging a relevant result | Combined *Extract Information* with *Reflect Stop* | As information is extracted from individual results, decision about its relevance is made. Content documents judged as relevant are saved by means of bookmarking and tag description. |

The interaction logs captured by Morae and by the secondary task program were imported into a MySQL database. The import process included an on-the-fly segmentation of the logged data. The segmentation procedure used a state machine in which transitions between states were determined by the URLs, the names of active program windows (in focus), and keyboard and mouse activity. The sequential stream of events was processed one event at a time. However, to detect text finding on a web page (that is, Ctr-F keystroke sequences), a ten-event history window was used. Pseudo-code for a simplified version of the algorithm is presented in Appendix B (Figure 7). The results of the automatic task stage segmentation were verified with the screen cam recordings and log files; and manually corrected as needed. The difference between the results from the algorithm and the algorithm+manual correction measured as a difference in labeling individual task stages was 5.05% (calculated as the mean of differences weighted by the number of instances of task stages). The largest difference was for the Q:Query (+9.2%) and H:Search engine home (-9.7%) stages. This difference was due to incorrect recognition of keystroke patterns by the algorithm.

## Results

We first present statistics that describe observed task stages. We omit stage M (subsequent result pages) from further considerations due to its low frequency (observed occurrences were 15 out of 288 tasks, i.e. 5% of cases). This corroborates results from other studies of web search engine use where users were reported to rarely move beyond the first results page. Selected descriptive statistics for task stages are presented in Table 9. Figure 5 depicts the most frequent transitions along with observed transition probabilities.

Table 9. Task stage descriptive statistics.

| Task stages | | Max per task | Mean per task | SD | Total for all tasks in the data set |
|---|---|---|---|---|---|
| All | all | 99 | 23.8 | 17.58 | 6852 |
| | unique | 63 | 17 | 11.02 | 4899 |
| C | all | 60 | 9.7 | 9.1 | 2778 |
| | unique | 34 | 5.92 | 4.99 | 1699 |
| L | all | 33 | 6.53 | 6.07 | 1880 |
| | unique | 17 | 3.76 | 3.18 | 1082 |
| B | | 16 | 2.88 | 2.08 | 803 |

| Task stages | Max per task | Mean per task | SD | Total for all tasks in the data set |
|---|---|---|---|---|
| Q | 13 | 3.56 | 2.88 | 1025 |

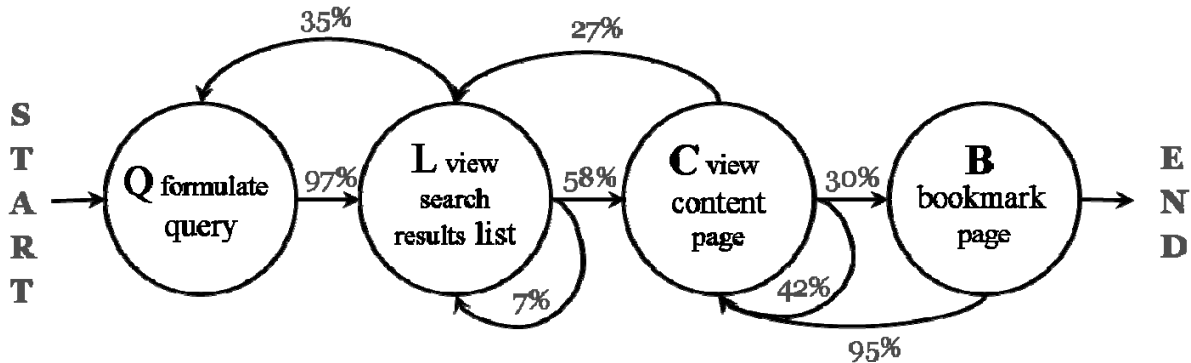

Figure 5. Task stage transitions with the most frequent transition probabilities (>5%) observed in the study.

## Components of cognitive load

Reaction time can be considered as containing a component related to person (e.g., dependent on motor, perceptual, and cognitive ability of the person), a component related to task stage (dependent on the cognitive demands of the task stage), and some other components (Equation 1).

$$RT_{total} = RT_{person} + RT_{task\_stage} + C \qquad (1)$$

## Average cognitive load per task stage

We calculated two secondary-task performance measures: a) the rate of missed secondary task events and b) reaction time to the secondary task. In 70% of search tasks no secondary tasks were missed; thus the variations in the level of miss rate represent only a relatively small part of the search tasks. We concluded that the miss rate in our study was a less sensitive measure than the reaction time. Therefore, our assessment of cognitive load focused on use of reaction time.

We conducted univariate analyses of variance (ANOVA) with objective task difficulty, task stage, user interface and cognitive abilities as fixed factors and the mean reaction time[6] per task stage as a dependent variable. The analyses were performed for the total reaction time ($RT_{total,\_all}$) as well as for color-match and no-color-match components ($RT_{total\_match}$ and $RT_{total,\_nomatch}$).

For $RT_{total,\ all}$ the analysis revealed significant differences among task stages and both cognitive abilities (Table 10). There was no significant effect associated with the objective task difficulty or the user interface. Post-hoc analysis (Tukey HSD test) showed the mean reaction times during the query stage Q were (borderline) significantly longer than during the search results list stage (L) (p=.058) and significantly longer than during the viewing content stage (C) (p=.003). Mean reaction times during the bookmarking stage (B) were significantly longer than during the viewing content stage (C) (p=.03). Other differences among the stages were not significant. The reaction times for the task stages are shown in Table 11.

---

[6] Reaction time variables were transformed by using logarithmic function to obtain a more symmetrical normal distributions.

Although we reported the measure of miss rate was overall less sensitive than reaction time, we also checked the effect of task stage on the overall miss rate. The effect of task stage on miss rate was significant (F(3,900)=3.53, p<.05). Post-hoc tests (Tukey HSD) showed that miss rate for task stage B was higher than for C (p=.029) and (borderline) higher than for L (p=.071). The direction of this effect is the same as for the reaction time.

Table 10. Significant effects (ANOVA) of factors on the average reaction times per task stage.

| Reaction time variables | | Factors | | | | | |
|---|---|---|---|---|---|---|---|
| | | Task Stage | User Interface | Task Stage * User Interface | Obj. Task Difficulty | Working Memory | Mental Rotation |
| **Average RT: total** | all | F(3,839)=5.97; p<.001 | | | | F(1,839)=6.8; p<.01 | F(1,839)=4.1; p<.05 |
| | match | | F(1,487)=4.6, p<.05 | F(3,487)=3.3; p<.05 | | | F(1,487)=4.9; p<.05 |
| | nomatch | F(3,767)=4.97; p<.005 | F(1,767)=5.2; p<.05 | F(3,767)=2.73; p<.05 | F(2,767)=3.0; p=.05 | F(1,767)=5.9; p<.05 | F(1,767)=3.7; p<.053[+] |

[+] Borderline significance (.05<p<.099)

For $RT_{total\_match}$ and $RT_{total\_nomatch}$, we found effects of user interface as well an interaction effect of user interface and task stage. We will discuss these effects in detail in the subsection on the user interface effects. For both $RT_{total\_match}$ and $RT_{total\_nomatch}$, we found significant effects of cognitive abilities. Perhaps more interestingly, we found a significant effect of task stage and of objective task difficulty on $RT_{total\_nomatch}$, but not on $RT_{total\_match}$ (Table 10). Post hoc tests revealed that the load on easiest tasks was significantly lower than on medium-difficulty tasks. User interface effects will be discussed in a separate section.

Table 11. Average reaction times across task stages (mean values and std. error)

| Reaction time [ms] | | Overall | | Task stage | | | |
|---|---|---|---|---|---|---|---|
| | | | | Q | L | C | B |
| **Average RT: total** | all | 2456 (32) | F(3,839)=5.97; p<.001 | 2628 (85) | 2373 (69) | 2266 (47) | 2488 (58) |
| | match | 2380 (44) | not sig. | 2591 (137) | 2316 (92) | 2294 (67) | 2421 (90) |
| | no-match | 2441 (34) | F(3,767)=4.97; p<.005 | 2588 (93) | 2395 (76) | 2243 (49) | 2475 (62) |

Task stages that we found to be characterized by slower secondary task reaction time involved more keyboard activity as the searchers were typing queries or tags. To confirm that the effects of task stages on slower reaction time could be interpreted as a higher cognitive load we checked whether the amount of keyboard activity (number of keyboard presses and the total time spent on keyboard interaction) was correlated with the reaction times to the secondary task for task stages B and Q. We found no significant correlations. Hence, the slower reaction times can be interpreted as mainly due to cognitive processes and not to motor activity (typing and switching one hand between a keyboard and a mouse).

## Individual Variability and Residuals

The two potentially significant sources of variability in reaction time, individual user characteristics and task stage, were considered separately. An analysis of variance was performed with participant identifiers as the main factor and reaction time as the dependent variable. The predicted mean reaction times and the residuals were then analyzed in separate analyses of variance as dependents with objective task difficulty, task stage, user interface and cognitive abilities as independent factors. This statistical procedure is equivalent to separating the typical mean reaction time for a user in the given circumstances (represented by $RT_{person}$) from the total reaction time (Madrid, van Oostendorp & Melguizo, 2009). This procedure should remove variability in the resulting reaction time (the residual value that is represented by $RT_{task\_stage}+C$) due to the differences among individual participants (including the differences in motor and cognitive skills). This was confirmed by the finding that the users' reaction time ($RT_{person}$) was significantly related to the users' cognitive abilities, while the residual reaction time ($RT_{task\_stage}+C$) was significantly related to the task stage, but not to their cognitive abilities (Table 12). Clearly, the differences between task stages were reflected in different reaction times to the secondary task events.

Table 12. Significant effects (ANOVA) of factors on the average reaction times per task stage.

| Reaction time variables | | Task Stage | User Interface | Task Stage * User Interface | Working Memory | Mental Rotation |
|---|---|---|---|---|---|---|
| **Average RT: person** | all | | | | $F(1,839)=34.8$, $p<.001$ | $F(1,839)=3.85$, $p=.05$ |
| | match | | | | | $F(1,481)=9.98$, $p<.005$ |
| | nomatch | | | | $F(1,761)=43.38$, $p<.001$ | $F(1,761)=3.54$, $p<.06^+$ |
| **Average RT: residual** | all | $F(3,839)=9.64$, $p<.001$ | $F(1,839)=3.1$, $p=.079^+$ | $F(3,839)=2.27$, $p=.079^+$ | | |
| | match | $F(3,481)=2.67$, $p<.05$ | $F(1,481)=3.92$, $p<.05$ | | | |
| | nomatch | $F(3,761)=2.15$, $p=.093^+$ | $F(1,761)=4.47$, $p<.05$ | $F(3,761)=2.79$, $p=.04$ | | |

[+] Borderline significance ($.05<p<.099$)

For $RT_{residual\_all}$, post hoc analysis (Tukey HSD) revealed that the mean reaction times during the query stage Q were significantly longer than during the search results list stage L ($p=.007$) and significantly longer than during the viewing content stage C ($p<.001$). Mean reaction times during the bookmarking stage B were significantly longer than during the viewing content stage C ($p=.001$). Other differences among the stages were not significant. Compared to the post-hoc analysis presented in section "Average cognitive load per task stage", the significant differences in reaction times $RT_{total\_all}$ between the stages Q, B, L, and C for $RT_{residual\_all}$ were in the same direction, but were more significant (Table 14).

Table 13. Summary of significant post-hoc effects (Tukey HSD) for RT total and residual.

| | Significance level (p-level) | |
|---|---|---|
| **Cognitive load effect** | $RT_{total\_all}$ | $RT_{residual\_all}$ |
| Higher cognitive load during Q than during L | $p=.058$ | $p=.007$ |

| Higher cognitive load during Q than during C | p=.003 | p<.001 |
| Higher cognitive load during B than during C | p=.030 | p=.001 |

## Peak cognitive load

The results show significant differences between the average levels of cognitive load across task stages. In addition to calculating average values of cognitive load, we calculated peak load, as maximum reaction time, for each task stage.

The total and the residual peak reaction times ($RT^{peak}_{total\ all}$, $RT^{peak}_{residual\ all}$) differed significantly for task stage and for objective task difficulty. The total and the residual peak reaction times for no-match cases ($RT^{peak}_{total\ nomatch}$, $RT^{peak}_{residual\ nomatch}$) differed for objective task difficulty (Table 14). Post-hoc tests (Tukey HSD) indicated that in all cases the shortest reaction time was for the easiest tasks (simple fact-finding) (p<.001), while there was no difference between difficult and medium difficult tasks.

For task stage, post-hoc tests showed that reaction time during viewing content pages stage (C) was significantly longer than during examining search result list stage (L) (p=.017) and borderline significantly longer than during the query formulation stage (Q) (p=.094). There was no significant difference between C and B stages. The difference between (C) and (L) is similar to the result obtained by Kim & Rieh (2005). The peak reaction times for the task stages are shown in Table 15.

Table 14. Statistics for the peak reaction times (Anova).

| | | **Factors** | | | | |
|---|---|---|---|---|---|---|
| **Reaction time variables** | | **Task Stage** | **User Interface** | **Obj. Task Difficulty** | **Working Memory** | **Mental Rotation** |
| **Peak RT: total** | all | F(3,839)=3.82, p=.01 | | F(2,839)=11.6, p<.001 | | |
| | match | | | | | F(1,487)=5.34, p<.05 |
| | no-match | | | F(2,767)=9.7, p<.001 | F(1,767)=6.2, p<.05 | |
| **Peak RT: residual (const + task stage)** | all | F(3,839)=2.65, p<.05 | | F(2,839)=10, p<.001 | | |
| | match | | | | | |
| | no-match | | | F(2,767)=8.33, p<.001 | | |

Table 15. Peak reaction times across task stages and user interface conditions. (mean and std. error)

| | | **Task stage** | | | | | |
|---|---|---|---|---|---|---|---|
| **DT reaction time [ms]** | | **Overall** | | **Q** | **L** | **C** | **B** |
| **Peak RT: total** | all | 3422 (55) | sig | 3361 (148) | 3285 (120) | 3648 (103) | 3367 (98) |
| | match | 2747 (58) | not-sig | 2786 (173) | 2595 (122) | 2849 (101) | 2728 (107) |
| | no-match | 3244 (55) | not-sig | 3077 (128) | 3192 (124) | 3434 (102) | 3175 (101) |

## User interface

We examined the nature of user interface main effects, as well as of interaction effects of user interface and task stage on $RT_{all}$, $RT_{match}$ and $RT_{nomatch}$. We found significant effects for total RT as well as for residual RT. The main effects are reported in Table 10 and Table 12. The interaction effects between UI and individual task-stages are reported in Table 16. Interestingly, we found that the direction of effects for color match and non-match cases differed. For color match, the average reaction time to the secondary task tended to be shorter in the UI1 condition (Google) than in the UI2 condition (ALVIS). For non-color match, the reverse was true (Table 17, Table 18, and Figure 6). Closer examination at the task-stage level revealed that the interaction effects of user interface and task stage were for color match cases due to differences in reaction time on task stage L and C (for C only for $RT_{total\_match}$), while in the non-color match cases the interaction effects were due to differences task stage Q and L (Table 18).

Table 16. Statistics for significant post hoc tests of interaction effects of task stage and user interface conditions.

| Reaction time [ms] | | Significance of overall interaction effect (Table 10 & Table 12) | Task stage | | | | | |
|---|---|---|---|---|---|---|---|---|
| | | | Q | | L | | C | |
| | | | UI1 | UI2 | UI1 | UI2 | UI1 | UI2 |
| **RT: total** | all | not-sig | | | | | | |
| | match | sig | | | t(106)=3.3, p=.001 | | t(182)=-2.1, p<.05 | |
| | no-match | sig | t(146)=2.3, p<.05 | | t(176)=2.1, p<.05 | | | |
| **RT: residual** | all | | | | | | | |
| | match | not-sig | | | t(110)=-2.3, p<.05 | | | |
| | no-match | sig | t(163)=2.1, p<.05 | | t(180)=2.7, p<.01 | | | |

Table 17. Average reaction times across user interface conditions (mean and std. error)

| Reaction time [ms] | | | User Interface | |
|---|---|---|---|---|
| | | | UI-1 (G) | UI-2 (A) |
| **Average RT: total** | all | not-sig | 2496 (49) | 2420 (42) |
| | match | sig | 2277 (64) | 2464 (61) |
| | no-match | sig | 2503 (53) | 2386 (45) |

Table 18. Average reaction times for significant post-hoc tests of interaction effects of task stage and user interface conditions (mean and std. error)

| Reaction time [ms] | | Task stage | | | | | |
|---|---|---|---|---|---|---|---|
| | | Q | | L | | C | |
| | | UI1 | UI2 | UI1 | UI2 | UI1 | UI2 |
| **RT: total** | match | | | 1985 (118) | 2528 (124) | 2157 (90) | 2420 (91) |
| | no-match | 2843 (162) | 2437 (111) | 2632 (147) | 2222 (74) | | |

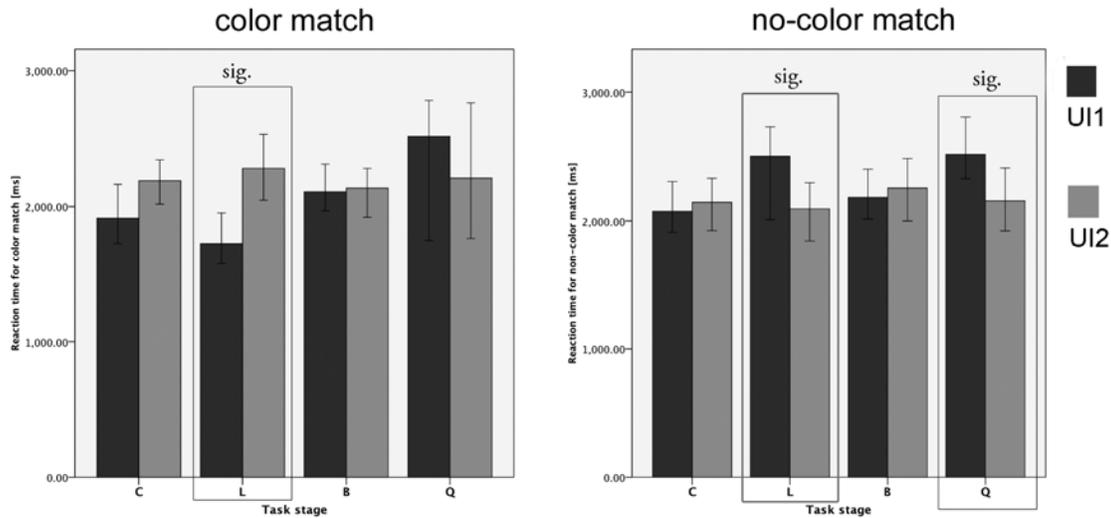

Figure 6. Interaction effects of task stages and user interfaces on reaction times in color match and non-color match variations of the secondary task (milliseconds, mean and std. error). Significant differences between the interfaces are marked by the rectangles.

## Other factors and the components of cognitive load

A person's knowledge and interest have been shown to influence cognitive load and task performance, for example, in effects of knowledge on mental workload (Xie & Salvendy (2000), and on information search strategies (Wildemuth, 2004; Zhang, Anghelescu & Yuan ,2005). In this section we analyze participant self-ratings of their *task familiarity* and *task interest*. There were no significant effects of *familiarity* and *interest* on the residual RT (i.e., on $RT_{task\_stage}+C$). This echoes the results for cognitive abilities. Familiarity effects were significant only for $RT_{person}$ in the total case and the *no-color-match* case. For the post-hoc results, we found that a significant difference appeared only in one case: for *no-color match* between *no familiarity* (1) and *medium familiarity* (3). The reaction times for *no familiarity* were faster than for *medium familiarity*. The effects of interest were significant for RT and for $RT_{person}$ and the effects disappeared for residual RT. Post-hoc tests (Tukey HSD) showed no significant effects for the total RT. In all other cases when there was a significant effect on RT, the post-hoc tests demonstrated that the significant difference was due to faster reaction times only in the *no interest* cases vs. all other cases. Of the 288 (total) tasks, 42 were rated by participants as no interest. After removing these cases from the data set, the significant effects on cognitive load remained (i.e., task stage cognitive load per task stage and task stage and task for the peak cognitive load). Some significance levels were affected with a few effects becoming somewhat less significant, while a few were more significant (e.g., for the total residual peak: peak reaction time during C-stages was longer than during L-stages at p=.05 and longer than during Q-stages, but at a borderline significant level of p=.07 ). We conclude that task stage effects on cognitive load were independent of user familiarity and user interest effects.

## Discussion

Our goal is to achieve a better understanding of mental demands imposed on users by search tasks and search interfaces. We found differences in cognitive load levels between search task stages. Specifically, as assessed by the average cognitive load on a task stage, query formulation (Q) and tagging of the relevant results (B) were associated with higher cognitive load than viewing results lists (L) and viewing individual content documents (C). This finding confirmed our first expectation. The Q and B stages were associated with the production of words by the searchers: query terms in the first case and descriptors (tags) of a bookmarked relevant result in the second. In these stages, additional working memory "slots" were needed for recalling the terms from long-term memory and possibly for holding in working memory information from previously seen documents (Rouet, 2006). In task stages L and C, cognitive processing involved more recognition than recall, and so these stages were easier for searchers.

A second, slightly different, interpretation of our results is based on the role of executive attention control as a mediator of cognitive capacity and a constraint on effective cognitive capacity of a person (Engle, 2002). For the measures of average cognitive load within a task stage Q and B were found to impose more mental demands. In these phases, searchers produced query terms or relevant document descriptors (tags). The increased mental demand can be explained by directing attention by the searchers to the internal mental processing, that is, to language production.

Measures of peak cognitive load showed that viewing content documents (C) imposed more mental demand than examining search result lists (L). Peak load on Q and B was statistically indistinguishable from C, the higher mental load on Q and B was explained earlier and here we will explain the high peak load on C by comparing it with L. During stage L, searchers examine results lists in order to assess each result's relevance to the current information need, while during stage C, searchers visit individual results in order to assess content relevance to the current information need. In both these stages, the focus of the searcher's mental processing is on an external information object and processing of the object involves skimming or reading text and viewing images[7]. The low average load on C and L indicates relatively low mental effort invested by the searchers in skimming or reading text. Searchers might have become skilled at quick processing of web pages and at expert skimming. However, in contrast to the results lists in L, some content web pages in C contained longer text passages and therefore required reading of at least some longer text. Plausibly, the peak cognitive load reflected these longest episodes of focused attention and reading.

We found a significant effect of task stage and of objective task difficulty for no-color-match secondary task cases, but there was no such effect for color-match cases. The result seems to indicate that the color-match secondary task condition may be sensitive to extraneous load (user interface), while the no-color-match secondary task condition may be sensitive to extraneous load (user interface) and both intrinsic load (task). The significant effects of cognitive abilities plausibly support this result. The color-match case was affected by mental rotation, while the no-color-match case was affected by working memory and mental rotation. If we consider mental rotation as related to visual-spatial processing involved in interface interactions, and working memory as related to semantic processing needed to process search task information and text in user interface, the obtained pattern is similar to the user interface and objective task effects.

---

[7] Given the document corpus used in the study from Wikipedia, we know that the content pages contained mostly text.

Considering the effect of search interfaces, we found that for the color match and no-color match cases, search interfaces affected cognitive load in opposite ways. For color match, higher cognitive load was associated with the Alvis interface. For no-color match the opposite was true. Further examination of the effect of interaction between the search interface and task stage, revealed the difference in color-match cases was due to significantly longer reaction times in Alvis compared with Google during task stage L. In contrast, for no-color match cases it was due to significantly longer times in Google compared with Alvis on task stages Q and L. Google (UI1) presents results in a simple list (Figure 1). Alvis (UI2) added categories to the results list (Figure 2). As expected, the increased complexity of interactions with search results list in Alvis imposed a higher mental effort on task stages L. That effect was found in the color-match secondary task condition. According to our analysis of objective task difficulty, this effect can be considered as mainly due to extraneous load imposed by the interface. In the no-color-match condition, the situation was different. Here the secondary task imposed higher demands on semantic processing (Kane & Engle, 2003). The categories in the Alvis interface provided additional semantic information to searchers. The benefit from providing this information seemed to have outweighed the disadvantages from the increased interaction complexity, and the difference between the interfaces was reversed in stage L in the no-color-match condition. The observed significant difference between the interfaces during the Q task stage for no-color match can be explained by noticing that the Alvis interface supported query reformulation by clicking on category terms. This operation resulted in narrowing down the search results and, since it relied on recognition, it was cognitively easier to perform than modification of a query by typing in terms. The weakness of this explanation lies in the fact that query reformulation by narrowing down was used rather infrequently (only in 5% of all queries). However, the category terms were always available on screen and could provide support to searchers in formulating queries even if the terms were not clicked on to actively narrow the search. The semantic features in the Alvis interface were found to generally compensate for the increased semantic difficulty in no-color match secondary task cases.

## Conclusions

The aim of this paper has been to present assessment of cognitive load on search tasks as a way to characterize web search in terms of cognitive load and so gain a better understanding of the user's experience of the web search process. We described a user study that employed a dual-task method based on the Stroop effect as a technique for assessing cognitive load on web search tasks.

We found average cognitive load was found to vary by search task stages. It was significantly higher during query formulation and tagging of a relevant document as compared to examining search results and viewing individual documents. The more complex display of search results in Alvis was found to impose a higher cognitive load than Google. However, semantic information added to the search results lists in the form of result categories in Alvis interface was found to decrease mental demands during query formulation and examination of the search results list. To our knowledge, only three prior studies in information science (Dennis et al., 1998; Dennis et al., 2002; Kim & Rieh; 2005) addressed the question of cognitive load differences between search task stages. However, in each case, mental effort measurement was not the main focus.

The article contributes to a better understanding of cognitive load measurement by showing that measures of average cognitive load tend to be sensitive to dynamic changes in task demands, such as the changes between task stages, even when they are not sensitive to the differences at a higher granularity level, such as the changes between tasks. This finding explains why Schmutz et al. (2009) and our earlier analysis, (Gwizdka, 2009) did not find significant relationships between secondary task reaction times averaged at the task level.

Our secondary-task method employed a Stroop-like task (Stroop, 1935). The effects of the Stroop task are well known, but the task is typically performed by itself and not in the context of human-computer interaction. We employed this dual-task method as a technique for assessing cognitive load on web search tasks. The results demonstrate promise for use of this kind of secondary task assessment to separate measurement of the extraneous cognitive load (mainly due to a user interface and a system) and the intrinsic load (mainly due to a task itself). The method, however, does not address a separate measurement of the germane load.

The article also presents a method for classifying web search tasks into task stages that is based on semi-automatic analysis of user interaction logs with about 95% accuracy. A limitation of the method is that the classification rules were derived from URL patterns obtained for the studied search engines (Google, Alvis) and content web sites (Wikipedia). We note, however, that the rules could be modified to cover other cases.

A relatively large proportion of easy tasks, as assessed by participants, is a study limitation. In over 50% of cases, tasks were assessed as easy and only 15% of tasks as difficult. Although care was taken to vary the search task difficulty, the tasks may have been too easy for the study population, who are almost constantly online. Additionally, this population may be used to dealing with various issues in web search. That kind of experience could have skewed their assessment of web search task difficulty. Experimental tasks can be made more difficult by designing them to be more complex or by selecting participants from an appropriate population (e.g., older, or less experienced people). Although task difficulty is an issue, we note that the main focus of this paper on task stage variables mitigates this limitation. It is possible that by employing tasks with a wider range of difficulty levels one could more easily observe differences in the average cognitive load at the task level. One virtue of measuring cognitive load at lower levels of granularity is that differences can be measured even in cases where task difficulty is not well controlled.

The difference in the participants' familiarity with the two search interfaces is another limitation of the current study. The user interface effects need to be interpreted with caution. We found, however, that in certain conditions (task-stages Q, L with more semantic processing demanded by the secondary task) cognitive load imposed by the Alvis interface was lower than the load imposed by Google. Hence, the benefit from providing semantic categories seemed to outweigh the disadvantage of interacting with a somewhat less familiar and a somewhat more complex interface. This is at least a partial indication that training on the Alvis interface reduced the difference in familiarity.

Understanding mental effort imposed by task stages informs the design of search systems. It provides an indirect indicator of the likelihood a searcher may have "spare" mental capacity at a given point in the task process. When a user is not burdened with high cognitive load, they may be willing to provide additional information to the system (e.g., relevance feedback). It could also be used in the design of notification delivery from other computing tasks (Adamczyk & Bailey, 2004). We should point out that a high mental load measured while a user is engaged in a task is not necessarily a bad thing. As proposed in the Cognitive Load Theory, it could reflect the necessary intrinsic load and the good germane load (Sweller, van Merrienboer, & Paas, 1998). The described method of assessing cognitive load can be applied in other information search contexts. It could be used to corroborate results of evaluation frameworks such as the recently proposed approach by Wilson, schraefel & White (2009). Finer granularity measures of cognitive load that are calculated in real-time could be used to inform personalization and adaptation in user interfaces that are modified according to the user's mental effort. They could also be used in collaborative information retrieval systems, where human-human and human-computer co-operation is mediated by interactive software that "understands" the mental load of all its users. (For an example of such application see: Fan, Chen & Yen, 2010.)

Assessment of cognitive load at the task level, for both subjective and objective measures, does not reflect dynamic shifts in the demands on mental processing and thus cannot inform us when exactly during the search process users may encounter increased levels of cognitive load. Therefore, we suggest that (relatively) unobtrusive, dynamic, on-task assessment methods are more appropriate for understanding mental effort on information search. Such methods can be used together with traditional task-level measurements, for example subjective ratings and dynamic measurements calculated at the task-level. Understanding dynamic changes of cognitive load is important and useful for gaining a complete picture of the information task process. Only by looking at a sequence of changes in the mental load during task performance are we able to understand fully the cognitive demands of search tasks and search systems. This can be accomplished, for example, by assessing load imposed by task phases (as shown in this paper) and interactive elements of search systems using measurement methods that are sensitive to the dynamics of searchers' performance.

## Acknowledgements

tbd

# Appendix A

Table 19. Example search tasks used in the study (one for each combination of task type and structure).

| Type | Question text |
|------|---------------|
| FF-S | You love history and, in particular, you are interested in the Teutonic Order (Teutonic Knights). You have read about their period of power, and now you want to learn more about their decline. What year was the Order defeated in a battle by a Polish-Lithuanian army? |
| FF-H | A friend has just sent an email from an Internet café in the southern USA where she is on a hiking trip. She tells you that she has just stepped into an anthill of small red ants and has a large number of painful bites on her leg. She wants to know what species of ants they are likely to be, how dangerous they are and what she can do about the bites. What will you tell her? |
| FF-P | As a history buff, you have heard of the quiet revolution, the peaceful revolution and the velvet revolution. For a skill-testing question to win an iPod you have been asked how they differ from the April 19th revolution. |
| IG-H | You recently heard about the book "Fast Food Nation," and it has really influenced the way you think about your diet. You note in particular the amount and types of food additives contained in the things that you eat every day. Now you want to understand which food additives pose a risk to your physical health, and are likely to be listed on grocery store labels. |
| IG-P | Friends are planning to build a new house and have heard that using solar energy panels for heating can save a lot of money. Since they do not know anything about home heating and the issues involved, they have asked for your help. You are uncertain as well, and do some research to identify some issues that need to be considered in deciding between more conventional methods of home heating and solar panels. |

## Appendix B

```
Get next event
if URL event
    if sub-state == find_on_page
        sub-state = NULL    // URL event resets finding text on page
    elseif sub-state == backspace
        if URL  URL_list //new URL, last backspace was in query box
            state[previous] = Q:query formulation
    URL_list = URL_list.append(URL)
    if URL  {Google, Alvis, Wikipedia search}
        if  parseURLargs(URL, getResultPage)==1 // first results page
            state = L:search results list first results page
        else
            state = M:search results list beyond first results page
        query = parseURLargs(URL, getQuery)
    elseif URL  Wikipedia content page
        state = C:content page
    elseif URL Scuttle tagging system
        state = B:bookmarking
        tags  = parseURLargs(URL, getTags)
    else
        state = C:content page // other kind of content page
if window event
    if window title == "Find"  // skip text search on a page
        sub-state = find_on_page
if keyboard event
    if sub-state == find_on_page
        loop // skip next keyboard events until new URL
    if key == Ctrl-F // skip text search on a page
        sub-state = find_on_page
    elseif key == Backspace // backspace could be navigational
                            //       or part of a query entry
        sub-state = backspace
    else
        state = Q: query formulation
        // query will be extracted from search results URL that follows
```

Figure 7. Pseudo-code for the simplified task-segmentation algorithm.

# References

Adamczyk PD, & Bailey BP. (2004). If not now, when?: The effects of interruption at different moments within task execution. Proceedings of ACM Conference on Human Factors in Computing (CHI'2004).

Back, J. & Oppenheim, C. (2001). A model of cognitive load for IR: Implications for user relevance feedback interaction. Information Research. 6(2).

Baddeley, A. (1986). Working memory. Oxford: Clarendon Press.

Baddeley, A. (1999). Essentials of human memory. Psychology Press.

Bates, M. J. (1979). Information search tactics. Journal of the American Society for Information Science, 30(4), 205-214.

Beatty, J. (1982). Task-evoked pupillary responses, processing load, and the structure of processing resources. Psychological Bulletin, 91(2):276–92.

Belkin, N. J., Cool, C., Stein, A., & Thiel, U. (1995). Cases, scripts, and information-seeking strategies: On the design of interactive information retrieval systems. Expert Systems with Applications, 9(3), 379-395.

Bell, D. & Ruthven, I. (2004). Searcher's assessments of task complexity for web searching. In J. T. Sharon McDonald (Ed.), Springer.

Bhavnani, S. K. & Bates, M. J. (2002). Separating the knowledge layers: Cognitive analysis of search knowledge through hierarchical goal decompositions. Proceedings of the American Society for Information Science and Technology, 39(1), 204-213.

Borgman, C. L. (1986). Why are online catalogs hard to use? Lessons learned from information-retrieval studies. Journal of the American Society for Information Science, 37(6), 387-400.

Borgman, C. L. (1989). All users of information retrieval systems are not created equal: An exploration into individual differences. Information Processing & Management, 25(3), 237-251.

Brünken, R., Steinbacher, S., Plass, J. L., & Leutner, D. (2002). Assessment of Cognitive Load in Multimedia Learning Using Dual-Task Methodology. Experimental Psychology, 49(2), 109-119.

Brünken, R., Plass, J. L., & Leutner, D. (2003). Direct measurement of cognitive load in multimedia learning. Educational Psychologist, 38(1), 53-61.

Buntine W, Valtonen K, Taylor M. (2005). The ALVIS Document Model for a Semantic Search Engine. Proceedings of the 2nd Annual European Semantic Web Conference; May 29, 2005. Heraklion, Crete.

Byström, K. & Jarvelin, K. (1995). Task complexity affects information seeking and use. Information Processing and Management, 31(2), 191-213.

Campbell, D. J. (1988). Task complexity: A review and analysis. Academy of Management Review, 13(1), 40-52.

Card, S.K., Newell, A., Moran, T.P. (1983). The Psychology of Human-Computer Interaction, Lawrence Erlbaum Associates, Inc., Mahwah, NJ.

Cegarra, J. & Chevalier, A. (2008). The use of tholos software for combining measures of mental workload: Toward theoretical and methodological improvements. Behavior Research Methods, 40(4), 988-1000.

Chandler P, and Sweller J. (1991). Cognitive Load Theory and the Format of Instruction. Cognition and Instruction. 8(4): 293-332.

Cowan, N. (in press, 2009). The magical mystery four: How is working memory capacity limited, and why? Current Directions in Psychological Science.

Dennis, S., Bruza, P., & McArthur, R. (2002). Web searching: A process-oriented experimental study of three interactive search paradigms. Journal of the American Society for Information Science and Technology, 53(2), 120-133.

Dennis, S., McArthur, R., & Bruza, P.D. (1998). Searching the World Wide Web Made Easy? The cognitive load imposed by query refinement mechanisms. In: Proceedings of the third Australian document computing symposium (ADCS'98), Department of Computer Science, Univer- sity of Sydney, TR-518 (pp. 65–71).

Eagle. M. & Leiter, E. (1964). Recall and recognition in intentional and incidental learning. Journal of Experimental Psychology, 1964, 68, 58–63.

Engle, R. W. (2002). Working memory capacity as executive attention. Current Directions in Psychological Science, 11(1), 19-23.

Fan, X., Chen, P.C., & Yen, J. (2010). Learning hmm-based cognitive load models for supporting human-agent teamwork. Cognitive Systems Research, 11(1), 108-119.

Ford, N., Miller, D., & Moss, N. (2001). The role of individual differences in internet searching: An empirical study. Journal of the American Society for Information Science and Technology, 52(12), 1049-1066.

Ford, N., Miller, D., & Moss, N. (2005). Web search strategies and human individual differences: A combined analysis. Journal of the American Society for Information Science and Technology, 56(7), 757-764.

Gwizdka, J. (2009). Assessing cognitive load on web search tasks. The Ergonomics Open Journal, 2, 114 - 123.

Gwizdka J, Chignell M. (2004). Individual differences and task-based user interface evaluation: A case study of pending tasks in email. Interacting with Computers. 16(4): 769-797.

Gwizdka, J., Lopatovska, I. (2009). The role of subjective factors in the information search process. Journal of the American Society for Information Science & Technology, 60(12), 2452 - 2464.

Hancock, P.A. & Chignell, M.H. (1987). Adaptive Control in Human- Machine Systems. In P.A. Hancock (Ed.) Human Factors Psychology, Amsterdam: North-Holland.

Harper, S., Michailidou, E. & Stevens, R. (2009). Toward a definition of visual complexity as an implicit measure of cognitive load. ACM Transactions on Applied Perception (TAP). 6(2). 1-18.

Hart, S. G. & Staveland, L. (1988). Development of NASA- TLX (task load index): results of empirical and theoretical research," in Human Mental Workload, P. Hancock and N. Meshkati, Eds., pp. 239–250, North-Holland Press, Amsterdam, The Netherlands.

Hirshfield, L. M., Chauncey, K., Gulotta, R., Girouard, A., Solovey, E. T., Jacob, R. J. K., et al. (2009). Combining electroencephalograph and functional near infrared spectroscopy to explore users' mental workload. In Lecture notes in computer science. (pp. 239-47). Springer Berlin / Heidelberg.

Ingwersen, P. (1996). Cognitive perspectives of information retrieval interaction: Elements of a cognitive IR theory. Journal of Documentation, 52(1), 3-50.

Iqbal, S. T., Zheng, X. S., & Bailey, B. P. (2004). Task-Evoked pupillary response to mental workload in human-computer interaction. In CHI '04: CHI '04 extended abstracts on human factors in computing systems. ACM.

John, B. E. & Kieras, D. E. (1996). Using GOMS for user interface design and evaluation: Which technique?. ACM Transactions on Computer-Human Interaction (TOCHI), 3(4), 287-319.

Kane, M. J., & Engle, R. W. (2000). Working-memory capacity, proactive interference, and divided attention: Limits on long-term memory retrieval Journal of Experimental Psychology.Learning, Memory, and Cognition. 26(2), 336-358.

Kane, M. J., & Engle, R. W. (2003). Working-memory capacity and the control of attention: the contributions of goal neglect, response competition, and task set to Stroop interference. Journal of experimental psychology.General, 132(1), 47-70.

Kelly, D., Shah, C., Sugimoto, C. R., Bailey, E. W., Clemens, R. A., Irvine, A. K., et al. (2008). Effects of performance feedback on users' evaluations of an interactive IR system. In Iiix '08: Proceedings of the second international symposium on information interaction in context. ACM Press.

Kim, K.-S. (2001). Information seeking on the web: Effects of user and task variables. Library & Informatin Science Research, 23(3), 233-255.

Kim YM, Rieh SY. (2005). Dual-Task Performance as a Measure for Mental Effort in Library Searching and Web Searching. Proceedings of the 68th Annual Meeting of the American Society for Information Science & Technology. Oct. 28 – Nov. 2, 2005. Charlotte, NC.

Li, Y. & Belkin, N. J. (2008). A faceted approach to conceptualizing tasks in information seeking. Information Processing & Management, 44(6), 1822-1837.

Madrid IR, Van Oostendorp H, Puerta Melguizo MC. (2009). The effects of the number of links and navigation support on cognitive load and learning with hypertext: The mediating role of reading order. Computers in Human Behaviour. 25(1): 66-75.

Marchionini, G. (1995). Information seeking in electronic environments. Cambridge Series on Human Computer Interaction, 9

Michailidou, E. (2009, August). The user evaluation of the vicram visual complexity prediction algorithm. Vicram technical report 8. University of Manchester.

Miller, G. (1956). The magical number seven, plus or minus two: Some limits on our capacity for processing information. Psychological Review, 63(2), 81-97.

Moray, N. (1979). Mental Workload: Its Theory and Measurement. New York: Plenum.

Oberauer, K. (2002). Access to information in working memory: Exploring the focus of attention. Journal of Experimental Psychology: Learning, Memory, and Cognition., 28(3), 411-421. doi:10.1037/0278-7393.28.3.411

Norman, D. (1988). The Design of Everyday Things, Currency Doubleday.

O'Donnell, R.D & Eggemeier, F.T. (1986). Workload assessment methodology. In Boff, Kaufman, & Thomas (Eds.) Handbook of Perception and Human Performance, Vol. II, Cognitive Processes and Performance (New York: John Wiley). Ch. 42.

Polson, P. G., Lewis, C., Rieman, J., & Wharton, C. (1992). Cognitive walkthroughs: A method for theory-based evaluation of user interfaces. Int. J. Man-Mach. Stud., 36(5), 741-773.

Rosenholtz, R., Li, Y. & Nakano, L. (2007). Measuring visual clutter. Journal of Vision, 7(2), 1-22. Retrieved 1 March, 2009 from http://journalofvision.org/7/2/17/article.aspx

Rouet, J.F. (2006). The skills of document use: from text comprehension to Web-based learning. Mahwah, NJ: Lawrence Erlbaum Associates.

Schmutz, P., Heinz, S., Métrailler, Y. & Opwis, K. (2009). Cognitive Load in eCommerce Applications - Measurement and Effects on User Satisfaction. Adv in Human Comp Interact.

Schnotz, W. & Kürschner, C. (2007). A reconsideration of cognitive load theory. Educational Psychology Review, 19(4), 469-508.

Seng, J., Ko, I., & Lin, B. (2009). A generic construct based workload model for web search. Information Processing & Management. 45(5), 529-554.

Shrout, P.E., and J. L. Fleiss (1979). Intraclass correlations: Uses in assessing rater reliability. Psychological Bulletin (86): 420-428

Smith, E. E., & Jonides, J. (1997). Working memory: A view from neuroimaging. Cognitive Psychology, 33, 5–42.

Stroop JR. (1935). Studies of interference in serial verbal reactions. Journal of Experimental Psychology. 18(6): 643-662.

Sweller, J., van Merrienboer, J., & Paas, F. (1998). Cognitive architecture and instructional design. Educational Psychology Review, 10(3), 251-296.

Tsang, P. S. (2002). Mental workload. In W. Karwowski (Ed.), *International encyclopedia of ergonomics and human factors - 3 volume set*. (pp. 500 - 503). Routledge.

Teo, L. & John, B. E. (2008). Cogtool-Explorer: Towards a tool for predicting user interaction. In CHI '08: CHI '08 extended abstracts on human factors in computing systems. ACM.

Toms E, O'Brien H, Mackenzie T, et al. (2008). Task Effects on Interactive Search: The Query Factor. In: Fuhr N, Kamps J, Lalmas M, Trotman A. Eds. Focused Access to XML Documents. Lect Notes Comput Sci 4862. Springer Verlag. pp. 359-372.

Tungare, M. & Pérez-Quinones, M. (2009). Mental workload in multi-device personal information management. In CHI '09: Proceedings of the 27th international conference extended abstracts on human factors in computing systems. ACM.

van Gog, T., Kester, L., Nievelstein, F., Giesbers, B., & Paas, F. (2009). Uncovering cognitive processes:

Different techniques that can contribute to cognitive load research and instruction. Computers in Human Behavior, 25(2), 325-331.

Van Orden, K. F., Limbert, W., Makeig, S., & Jung, T. -P. (2001). Eye activity correlates of workload during a visuospatial memory task. Eye Activity Correlates of Workload During a Visuospatial Memory Task, 43(1), 111-121.

Wästlund, E., Norlander, T., & Archer, T. (2008). The effect of page layout on mental workload: A dual-task experiment. Computers in Human Behavior, 24(3), 1229-1245.

Wildemuth, B. M. (2004). The effects of domain knowledge on search tactic formulation. Journal of the American Society for Information Science and Technology, 55(3), 246-258.

Wilson ML, schraefel mc, White RW. (2009). Evaluating Advanced Search Interfaces using Established Information-Seeking Models. Journal of the American Society for Information Science & Technology. 60 (7): 1407-1422.

Wickens, C. D. (2002). Multiple resources and performance prediction. Theoretical Issues in Ergonomics Science. 3(2), 159-177.

Xie, B., & Salvendy, G. (2000). Prediction of Metal Workload in Single and Multiple Task Environments. International Journal of Cognitive Ergonomics. 4(3), 213-242.

Xie, H. (2002). Patterns between interactive intentions and information-seeking strategies. Information Processing and Management, 38(1), 55-77.

Zhang, X., Anghelescu, H.G.B. & Yuan, X. (2005) "Domain knowledge, search behaviour, and search effectiveness of engineering and science students: an exploratory study" Information Research, 10(2) paper 217 [Available at http://InformationR.net/ir/10-2/paper217.html]